# Controlled Spalling of Single Crystal 4H-SiC Bulk Substrates

*Connor P Horn[1,2], Christina Wicker[1,2], Antoni Wellisz[1], Cyrus Zeledon[1], Pavani Vamsi Krishna Nittala[1#], F Joseph Heremans[1,2], David D Awschalom[1,2], Supratik Guha[1,2]**

1 Pritzker School of Molecular Engineering, University of Chicago, Chicago, IL 60637 USA
2 Material Science Division and Center for Molecular Engineering, Argonne National Laboratory, Lemont, IL 60439 USA

* Please address correspondence and requests for materials to S. G. (email: guha@uchicago.edu)
# Now at Micron Technology, Inc., Boise, ID 83716 USA

Keywords: controlled spalling, 4H-SiC, power electronics, solid state qubits

We detail several scientific and engineering innovations which enable the controlled spalling of 10 – 50 micron thick films of single crystal 4H silicon carbide (4H-SiC) from bulk substrates. 4H-SiC's properties, including high thermal conductivity and a wide bandgap, make it an ideal candidate for high-temperature, high-voltage power electronic devices. Moreover, 4H-SiC has been shown to be an excellent host of solid-state atomic defect qubits for quantum computing and quantum networking. Because 4H-SiC single crystal substrates are expensive (due to long growth times and limited yield), techniques for removal and transfer of bulk-quality films in the tens-of-microns thickness range are highly desirable to allow for substrate reuse and integration of the separated films. In this work we utilize novel approaches for stressor layer thickness control and spalling crack initiation to demonstrate controlled spalling of 4H-SiC, the highest fracture toughness material spalled to date. Additionally, we demonstrate substrate re-use, bonding of the spalled films to carrier substrates, and explore the spin coherence of the spalled films. In preliminary studies we are able to achieve coherent spin control of neutral divacancy ($VV^0$) qubit ensembles and measure a quasi-bulk spin $T_2$ of 79.7 μs in such spalled films.

## 1. Introduction

Controlled spalling of semiconductors is a technique developed for removing thin (10 – 50 micron) layers from atop a semiconductor substrate by triggered and deliberate propagation of a sub-surface crack across the entirety of the chip or wafer.[1,2] Stress is built up in the wafer subsurface by the deposition of an appropriate metal (stressor) layer on the wafer surface. The crack originates at the wafer edge and then propagates laterally at a depth of 10 – 50 microns to relieve this stress without a need for post-conditioning (e.g. heat treatments). Spall depth can be modulated by engineering the stress field via the metal film deposition. A significant benefit of spalling is that the bulk-like properties of the exfoliated film are preserved[3–5] since the crack depth is determined by an elastic stress field, rather than an intervention by ion implantation or by the deposition of heterogeneous layers at the separation interface. The principal breakthrough in spalling was made by Bedell et al. who introduced a controllable method for spalling using nickel films deposited under high tensile stress via sputtering or electroplating.[1] This method has proven to be highly versatile, and to date has been used to spall Si, Ge, and III-V semiconductor wafers.[1,4–7] Silicon wafers of up to 300 mm in diameter have been spalled.[1] However, the materials spalled so far have been semiconductors with moderate to low fracture toughness and there have been no reports of successful spalling of more refractory, hard materials with a significantly higher fracture toughness.

One such material with a significantly higher fracture toughness is the technologically important semiconductor, silicon carbide (SiC), particularly the 4H polytype. 4H-SiC high-power electronics are being increasingly adopted in electric vehicles (4H-SiC MOSFET based inverters) and photovoltaic power management (4H-SiC high power diodes).[8,9] 4H-SiC is also a leading wafer scale candidate for solid state quantum coherent devices in quantum communications and sensing.[10,11] Successful spalling of 4H-SiC creates two principal unique opportunities. First, a hindrance to further widespread adoption of 4H-SiC is the high cost of manufacturing substrates. High intrinsic defect density, challenging polytype control, high temperatures, and long growth times contribute to low yields and high substrate cost.[12,13] Spalling offers a pathway for reusing a substrate multiple times if the spalled device layer can be integrated onto other substrates. Second, such layer removal via spalling motivates the heterogeneous integration of the resulting 4H-SiC membrane layers with other materials. This is particularly attractive for quantum technologies, where 4H-SiC has well characterized native defect-based qubits with long coherence times,[14] and these native defects may be located and spalled to be integrated with silicon-based control electronics or embedded on photonic waveguides for applications in quantum communication.[15]

By demonstrating successful spalling of 4H-SiC, we have overcome the challenge of spalling an ultrahard material which requires 2.5 times greater strain energy than needed to spall GaN, the previous hardest material to be spalled.[6] This result is enabled by novel scientific approaches taken in stressor layer design and spalling crack initiation (described in the Results and Discussion section). We present a controlled spalling-based solution for layer removal and transfer of few tens-of-microns thick films of single crystal 4H-SiC from bulk substrates. Bulk substrates are then repolished and can be reused to spawn further films for removal and transfer. We further show coherent spin control of a $VV^0$ qubit ensemble in 4H-SiC with $T_2$* and $T_2$ of the same order of magnitude as in bulk substrates.

## 2. Results and Discussion

Prior to spalling, a film of metal (typically Ni) is deposited onto the wafer to be spalled such that stresses in the metal layer give rise to an elastic stress field in the wafer sub-surface region.[16] The higher the fracture toughness of the wafer, the higher the thickness and stress of the metal film required to induce steady state spalling. The theoretical model which has proven to be valuable for predicting this thickness and stress is described by Suo and Hutchinson[17] and enables calculation of the stress intensity factors $K_I$ and $K_{II}$ of a propagating crack within the substrate. Details and an application of this model have been described by Bedell et al.[4] In summary, the crack originates at a free surface (usually the top surface of the semiconductor wafer) and propagates as a mixed mode crack (non-zero values of $K_I$ and $K_{II}$) plunging into the semiconductor. At a specific depth (predicted by the Suo and Hutchinson model) when $K_{II} \sim 0$, the crack propagates in a direction that is on average parallel to the surface and spalls off a film of the semiconductor substrate attached to the metal stressor layer.

Successful spalling, however, requires one additional condition which is demarcated by the critical strain energy release rate $G_C$ as described by Irwin and Orowan.[18,19] The spalling crack propagation must release more strain energy than the energy which is required to break the bonds in the crystal. $G_C$ is related to the material's intrinsic fracture toughness ($K_{IC}$) and is defined as

$$G_C = \frac{K_{IC}^2}{E\,(1-\nu^2)}. \quad (1)$$

Here, E is the Young's modulus and ν is the Poisson's ratio of the crack propagation medium. The semi-log Ashby plot in **Figure 1** illustrates the strain energy required for spalling 4H-SiC as compared to various materials which have previously been spalled. GaN was previously the most challenging material that had been spalled, whereas the 4H-SiC spalling demonstrated in this work requires almost 2.5 times more strain energy. The material properties used in this plot were gathered from a breadth of reported data[20–25] as well as the previous papers[4–6] on the spalling of each of the materials listed. Figure S1 gives further details on the required Ni metal stressor layers needed for spalling the selected materials. This previously unexplored level of strain energy per unit area applied to the metal-semiconductor spalling system poses new challenges critical to spalling of hard materials, notably (i) the thickness distribution of the metal film and (ii) the spalling crack nucleation. The scientific approaches used to address and overcome these challenges are described in this work and are expected to be applicable to the spalling of many other high fracture toughness semiconductors beyond 4H-SiC. We address below some of the key materials issues relevant to the spalling of such ultra-hard materials.

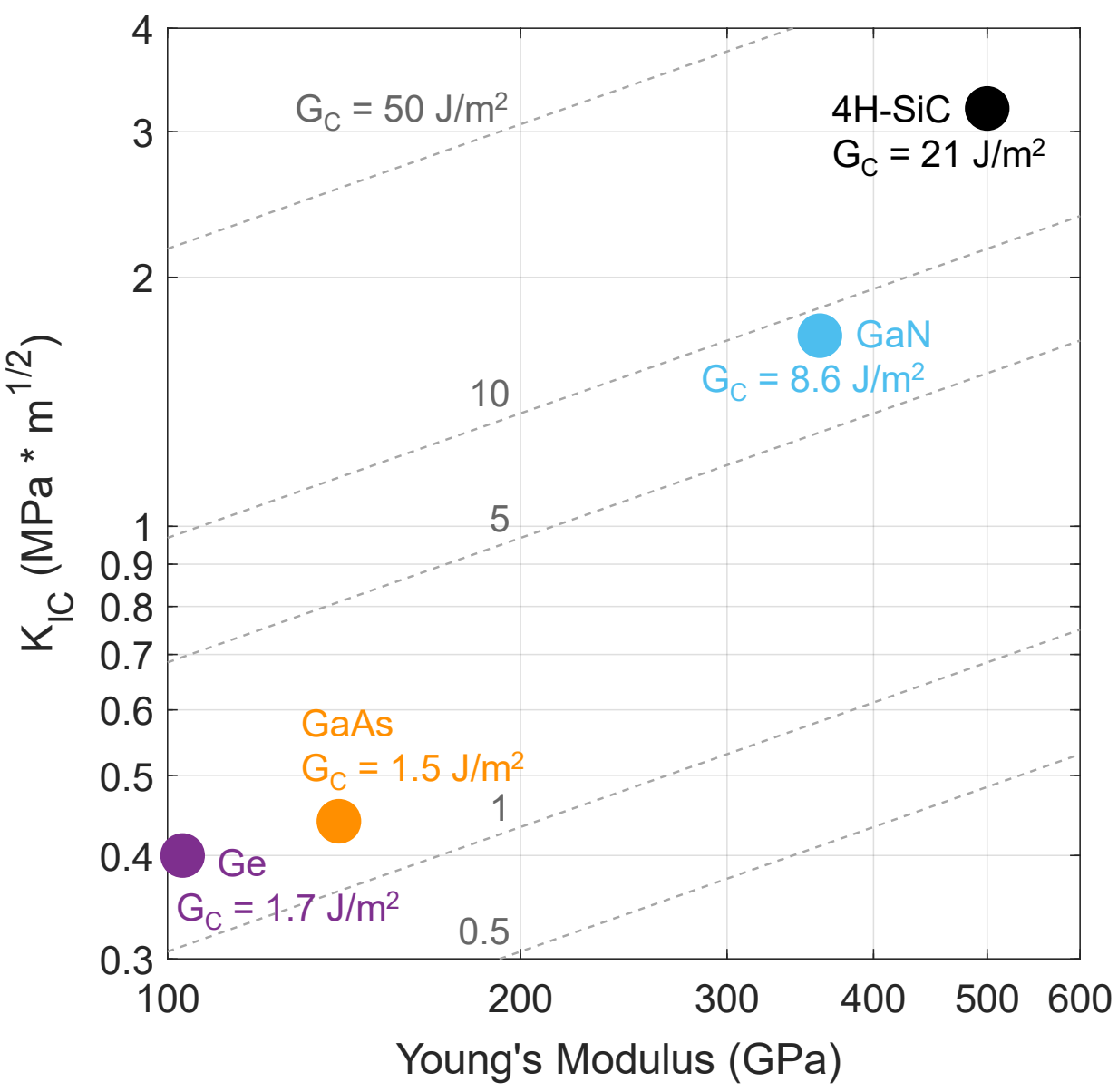

**Figure 1. Mechanical properties and strain energy comparison of spalled substrates.** The Ashby map depicts the mechanical properties of previously spalled substrate materials alongside 4H-SiC. The dashed lines denote equipotential $G_C$ energies, plotted using Equation 1. Note that the Poisson's ratio used for the equipotential lines was 0.25, an average value among these semiconductors.

### 2.1. (i) Thickness distribution of the metal film

The thickness distribution of the metal stressor layer holds considerable influence over the outcome of the controlled spalling process. Our studies have shown that the spalling fidelity of 4H-SiC is unpredictable if the thickness of the stressor layer at the substrate edges versus the substrate center varies by more than 10% from the intended distribution. This is because the fracture condition required for crack initiation and crack propagation are generally independent.[26] Electroplated nickel has been used extensively[5,6,27,28] as the metal of choice for spalling due to the availability of well characterized plating techniques, high deposition rate, and the ability to precisely control Ni stress through plating conditions.[29] The Ni is electroplated onto a thin Cr/Au or Ti/Au seed layer that is initially sputtered on the wafer (see methods section for details). Although thickness uniformity can be excellent for sputtered Ni as compared to electroplating, the deposition rate (~ 2 μm per hour) and maximum tensile stress (700 MPa)[4] of sputtered Ni are undesirable and unacceptable, respectively, to attain the > 20 μm thickness and ~ 700 – 850 MPa stress needed to spall 4H-SiC (see Figure S1). These stress levels have been demonstrated via electroplating of Ni,[6] but additional considerations are needed for ensuring desirable thickness distribution of the plated layer. Therefore, considerable effort was spent addressing a known aspect of electroplating called "current crowding" in which the deposit on the wafer edges can be up to 2-3 times thicker than the deposit thickness at the center.[30,31]

To address the problematic thickness nonuniformity of electroplated Ni, an auxiliary cathode which is known in the electroplating industry as a "thief"[32] was designed and integrated into the electroplating setup. The thief adds additional surface area to the cathode and can be used to control current density at the edges of the target substrate by altering the current distribution. **Figure 2a** shows a diagram of the electroplating bath with the thief shorted to and coaxially offset from the Au-coated 4H-SiC cathode (which, as note earlier, acts as the seed for electroplating Ni). An iterative design process for the thief utilized finite element modeling in COMSOL to simulate the Ni electroplating thickness profiles on a 29 x 29 mm square substrate (typical size of 4H-SiC die used in this work) as a function of the size, shape, and position of a conductive thief surrounding it. Details of the electroplating parameters used in these simulations are given in the methods section.

Iteratively tuning the thief position (offset) and the parametric curve that describes the shape of the thief allows us to independently adjust: (1) the Ni thickness at the absolute center versus the edges of the target 4H-SiC substrate (such that the edges can be at least 30% thicker or thinner than the center), and (2) the Ni thickness around the edges of the square substrate by a similar factor at any segment along the perimeter. This enables optimization of the thickness at the corners versus the sides. Figure 2b,c shows examples of the thickness variation control possible with and without the thief. Our findings allow us to ultimately choose the depicted thief shape and an offset such that the Ni thickness hierarchy is as follows: center of chip > edge centers > corners. This scheme was chosen subsequent to the observation that square substrates nearly always start spalling from the corners, so if the corners are the last regions to reach the critical Ni thickness required for spalling, the rest of the substrate will already have sufficient stress from the Ni to easily propagate the spalling crack.

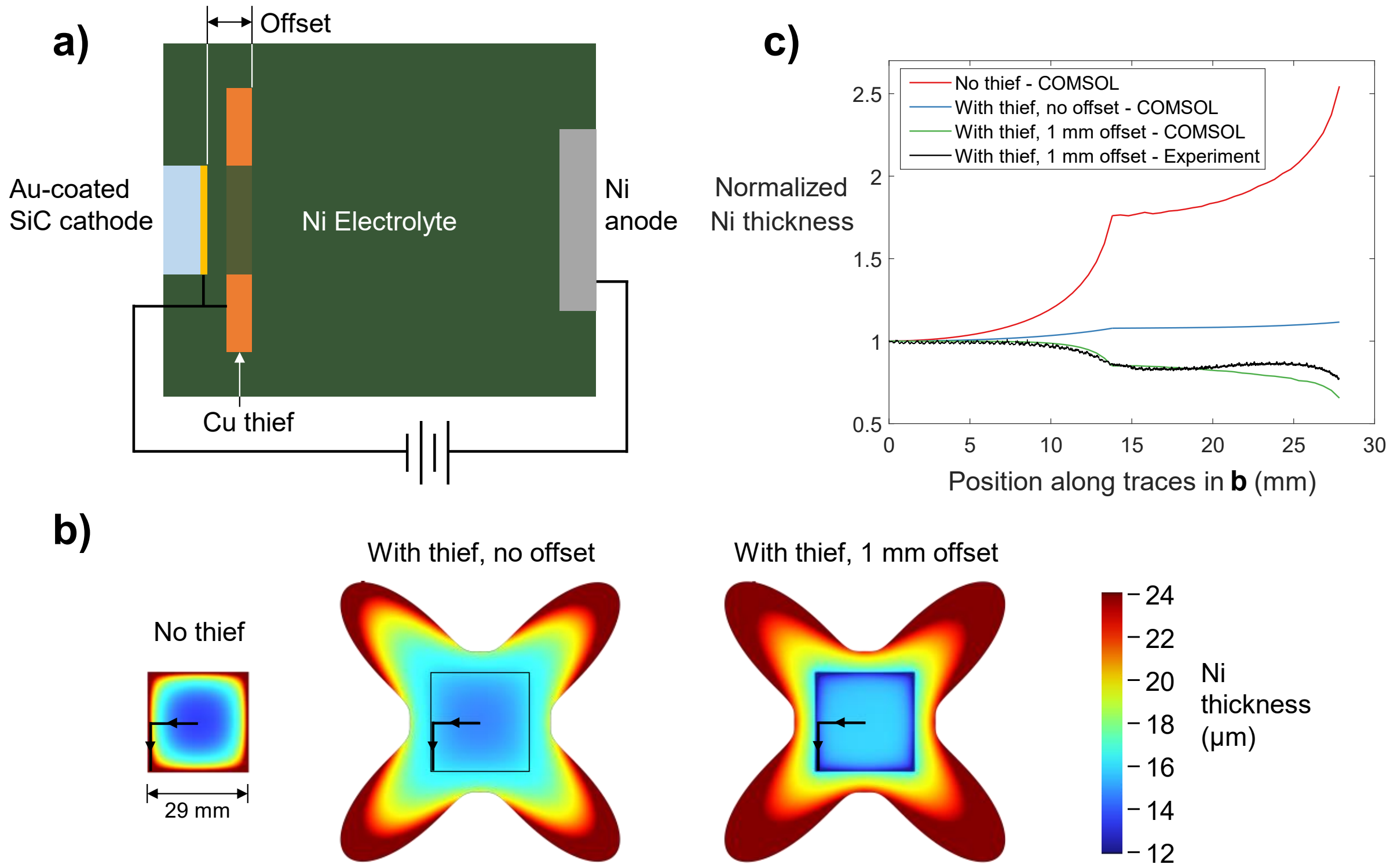

**Figure 2. Electroplating geometry and Ni thickness distribution. a)** Top-down schematic of the electroplating geometry. The opening in the thief exactly matches the square shape of the SiC cathode. **b)** COMSOL simulated Ni electroplating thickness distribution on 29 mm x 29 mm square substrates with and without thieves. Note that outer areas of the thieves exceed the plotted thickness range which is capped at 24 μm for better height resolution on the SiC surface. Current density averages 9 mA cm$^{-2}$ on the SiC surface in all three cases. **c)** Simulated and experimental measurements of Ni thickness along the traces in **b**. Values are normalized so that the thicknesses at the center of the substrates are coincident on the plot.

### 2.2. (ii) Spalling crack nucleation

Another concern for the spalling of ultrahard materials is that crack initiation – the prerequisite to steady state spalling – was found to be unreliable when utilizing controlled spalling techniques which have been established previously. The conventional method by which spalling crack initiation is made into Si and other semiconductors of similar toughness is to make the Ni abruptly discontinuous away from the substrate edge.[1,4] The stress concentration where the edge of the Ni meets the substrate is then high enough such that a crack can initiate either spontaneously when a certain Ni thickness is reached, or with external force from a handle layer of tape which is applied on top of the Ni and pulled upwards. In this case, no modification of the substrate is needed to initiate a spalling crack, i.e. $\theta = 0°$ as defined by the diagram in **Figure 3a**. For 4H-SiC however, we found that crack initiation fails when using this technique. Rather than inducing a spalling crack, the Ni delaminates from the Cr/Au or Ti/Au electroplating seed layer, specifically at the interface between the Au and Ni (see Figure S2).

The adhesion between the 4H-SiC substrate and the Cr or Ti was consistently observed to be strong and rarely a source of delamination. Moreover, the delamination of the Ni precisely at the Au-Ni interface was unexpected because the bond dissociation energy of the Au-Ni bond (247 kJ $mol^{-1}$) is very similar to the bond dissociation energy of the Au-Au bond (225 kJ $mol^{-1}$),[33] suggesting that this metal interface should not be a weak point. Our conclusion is that the delamination was likely triggered by the geometry of the layers rather than chemical bond strength, such that a modified geometry was needed to alter the local stress intensity where the edge of the Ni meets the 4H-SiC.

A solution to this crack initiation problem was found by modifying the 4H-SiC substrate such that $\theta > 0°$ as defined in Figure 3a. In practice, we chose to cut a shallow trench at the edge of the 4H-SiC substrate with a standard dicing saw after seed layer deposition, following which the Ni electroplating was carried out. By adjusting the cut depth, we were able to vary $\theta$ angles from 8° - 90° and initiate spalling at the top surface of the 4H-SiC substrate directly adjacent to the trench. Figure 3b,c depicts 3D and 2D laser confocal microscope scanned surface profiles at the crack initiation edge of a spalled 4H-SiC substrate with $\theta \sim 29°$. The trench spans from x = 0 to 0.15 mm, and then for x > 0.15 mm the spalling crack initiates from the 4H-SiC surface, plunging downward into the substrate until the equilibrium spall depth is reached. This method of inducing a spalling crack with an angled trench appears to be counter to the expectations of other relevant studies on crack initiation at the edge of stepped boundaries.[26,34] These suggest that as compared to larger $\theta$ angles, $\theta = 0°$ should concentrate the $K_I$ stress intensity most highly where the edge of the Ni meets the substrate and most strongly favor crack initiation. One possible explanation is that by removing material from the 4H-SiC adjacent to the edge of the Ni, the total strain energy needed to laterally separate the 4H-SiC substrate is decreased. Once the crack has been initiated, it is then free to propagate through the 4H-SiC substrate according to Suo and Hutchison model.[17]

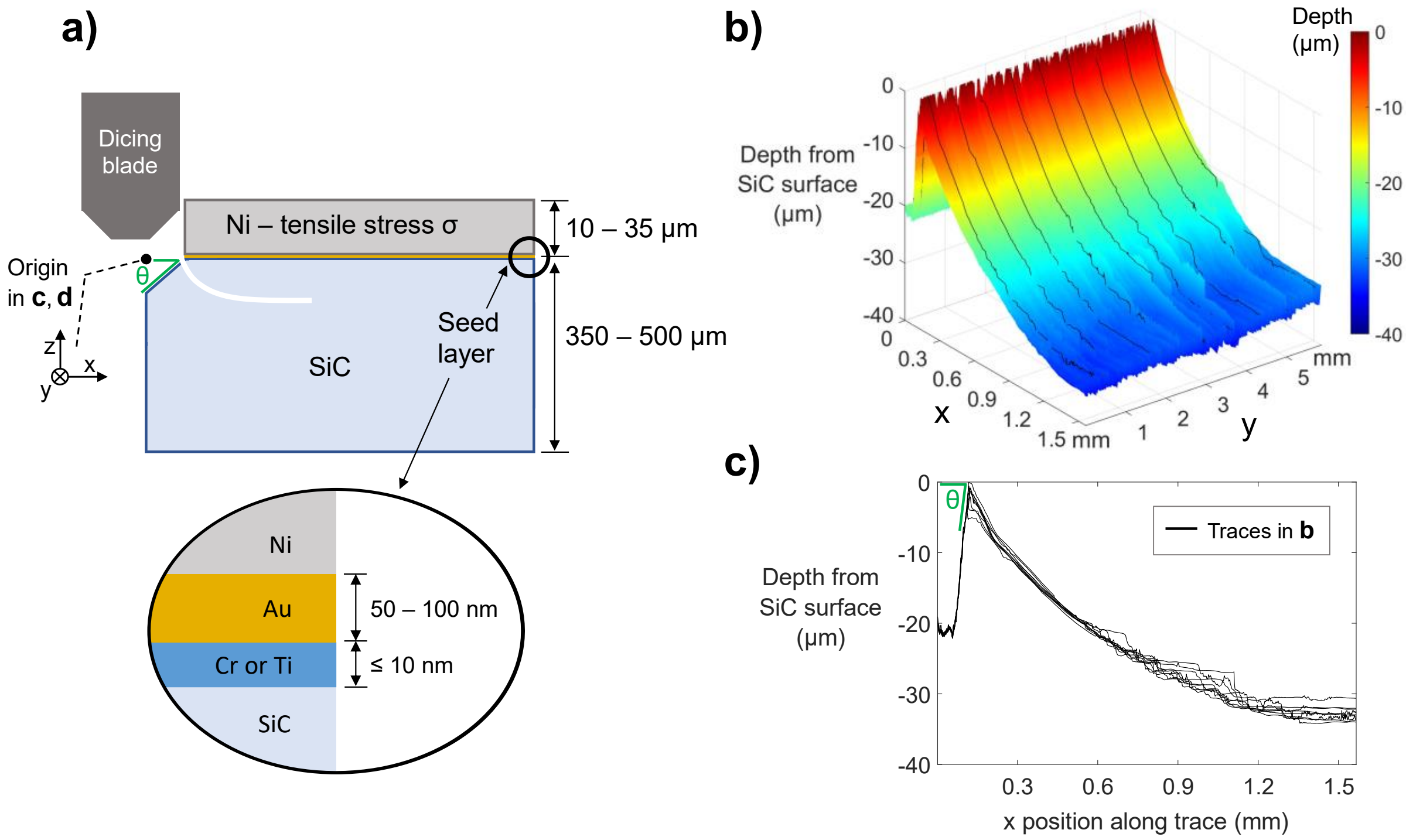

**Figure 3. Spalling crack initiation into 4H-SiC. a)** Diagram showing the spalling geometry for crack initiation and propagation. **b)** 3D laser scanning confocal microscope map taken at the crack initiation edge of a 4H-SiC substrate after spalling. Trench for crack initiation is present from x = 0 to 0.15 mm, beyond which the spalling crack is mapped. **c)** Black line traces on the 3D map in **b** are merged and plotted as spall depth versus x position. Trench angle θ is measured to be 29°.

### 2.3. Physical characterization of spalled substrates and films

The 4H-SiC substrates spalled in this work ranged from 5 x 5 mm to 29 x 29 mm squares cut from 100 mm or 150 mm wafers. The decision to spall square dies instead of full wafers reflects the high cost of 4H-SiC wafers and not any anticipated new challenges with spalling larger substrates. In **Figure 4**, 3D laser scanning confocal microscopy studies were carried out to further explore the surface morphology of the spalled face. A 5.5 x 5.5 mm area near the center of a spalled substrate is investigated in Figure 4a,b. The profiles reveal that spall depth along the direction of crack propagation ($[1\bar{1}00]$) is more variable (~ 6 µm peak-to-peak undulation) than spall depth perpendicular to the direction of crack propagation ($[11\bar{2}0]$) (< 1 µm fluctuation). This outcome can imply a slight variability in the speed or angle of crack propagation which in turn causes the crack depth to waver.[6] For applications which require a more homogenous spall thickness, a mechanical system can be adopted to better control the spalling crack speed and angle to improve this nonuniformity.[2]

4H-SiC substrates that were both on-axis (c-plane and m-plane) as well as off-axis (c-plane with 4-degree miscut towards $[11\bar{2}0]$) were used in this study, since the latter orientation is often used in power electronics applications. On-axis c-plane $(0001)$ and m-plane $(10\bar{1}0)$ 4H-SiC were both found to yield smooth corrugation-free surfaces as pictured in Figure 4c, establishing that both of these slip systems[20,35] exhibit favorable cleavage for spalling. When spalling the miscut substrates, matching corrugations were observed on the surfaces of the remaining substrates and spalled films, with amplitude and spall angle dependent on spall direction as plotted in Figure 4d. The perpendicular-to-miscut spall ($[1\bar{1}00]$ spall direction) had the lowest roughness: the mean ± standard error of the maximum peak-to-peak height measured across ten line profiles was 0.927 ± 0.015 μm. It is well known in spalling literature that semiconductors with ionic character (GaAs, InP) tend to spall along specific crystal planes, whereas elemental semiconductors are not as bound to this tendency.[1,2,5,7] SiC has comparable ionicity to GaAs (electronegativity difference 0.7 for SiC and 0.5 for GaAs),[36] which explains the sawtooth-like spalling of the miscut substrates.

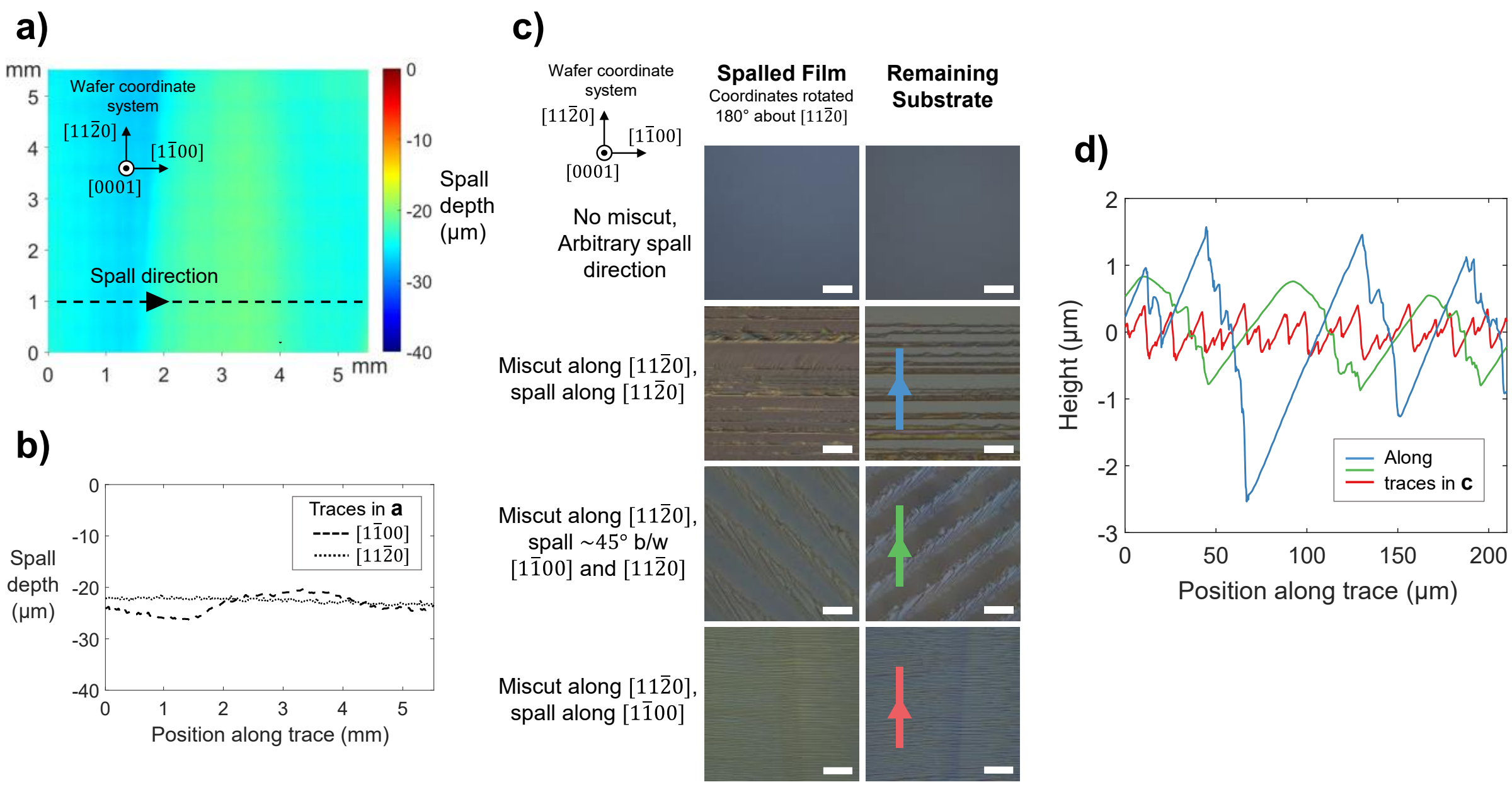

**Figure 4. Spalled 4H-SiC film and substrate surface morphology. a)** 3D laser scanning confocal microscope map of the surface of a remaining substrate after spalling. Absolute spall depth and tilt are calibrated by deliberately un-spalled regions outside the boundaries of the map shown. Wafer coordinate system shown assumes a non-miscut crystal. **b)** Profiles of spall depth versus position for the line traces in **a**. **c)** Optical differential interference contrast images of various spalled films and substrates. Blue, green, and red colored traces indicate the scan direction of the profiles in **d**. Wafer coordinate system shown assumes a non-miscut crystal. Scale bars, 100 μm. **d)** Laser scanning confocal microscope profiles of spall depth variation versus position for the line traces in **c**, centered about the mean surface heights.

### 2.4. Substrate re-use and integration of spalled films on carrier substrates

To demonstrate substrate re-use, a selection of previously spalled substrates was repolished using a standard lapping and chemical mechanical polishing procedure for 4H-SiC (see methods). The substrates were then re-spalled to establish that there are no unforeseen problems caused during surface reconditioning which would otherwise prevent further spalling. In total, the thickness reduction of the initial 4H-SiC substrate due to spalling and re-polishing can be limited to the maximum spall depth + approximately 15 to 20 μm due to the lapping and polishing process. Images of the repolished and re-spalled substrates are included in Figure S3.

Once the metals for spalling are etched away, the freestanding spalled 4H-SiC (typically 10 – 50 μm thick) is still rigid enough to be handled easily (see **Figure 5**) for transfer and bonding to a handle substrate. For example, we routinely bond the spalled films to a silicon wafer using a 25 μm thick epoxy-based die attached film from AI Technology Inc., as pictured in Figure S3. This bonding does not require any intermediate manipulations or carriers to be realized; the Ni is simply wet etched away from the spalled film and the film then pressed onto the bonding tape and heated to 120 °C to cure the epoxy bond.

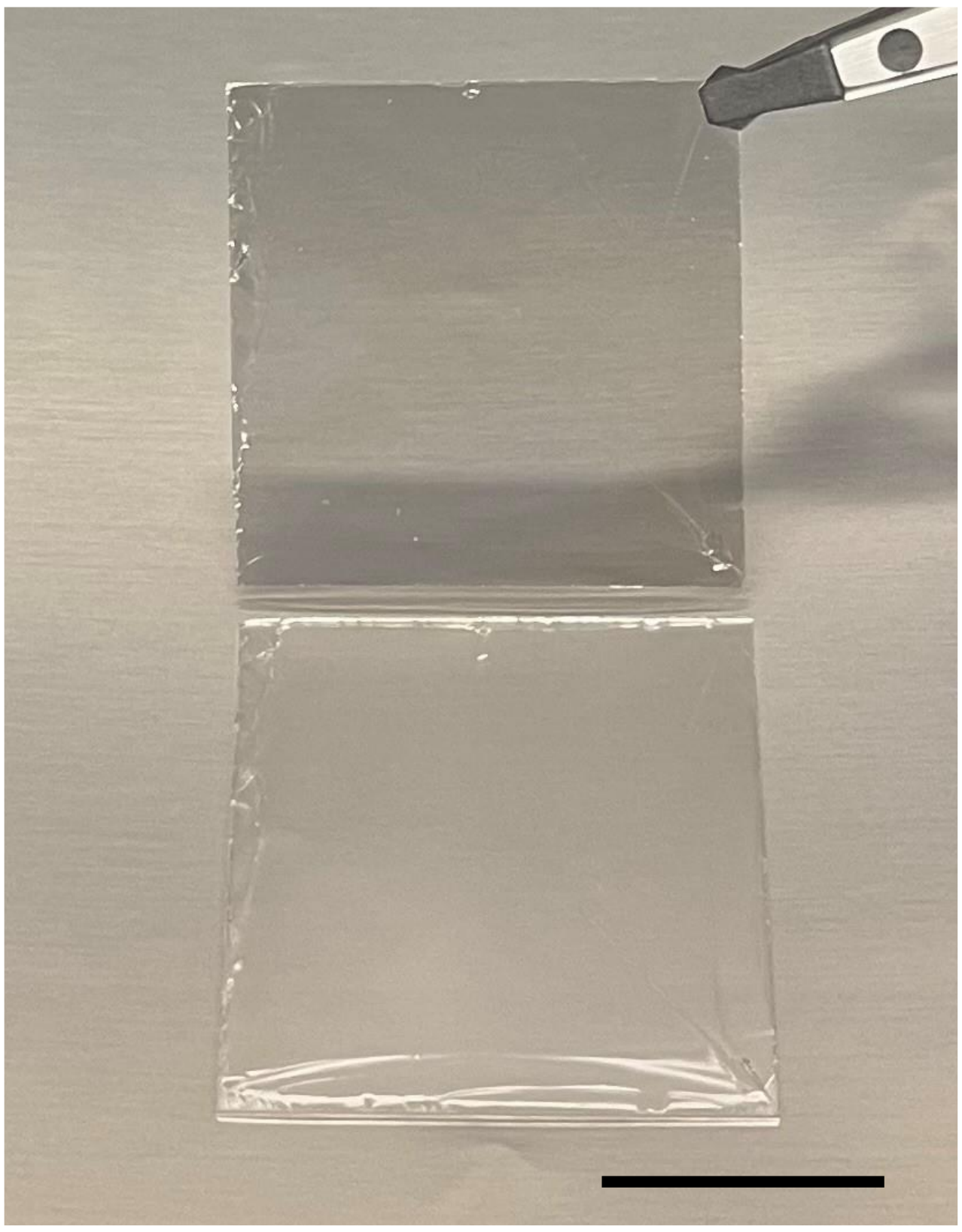

**Figure 5. Picture of spalled 4H-SiC and remaining substrate.** The ~ 30 μm thick spalled film is held with a tweezer above the corresponding substrate which it originated from.

**2.5. Measurement of qubit properties in spalled films**

As noted in the introduction, spalled 4H-SiC is an attractive approach for heterogeneously integrating 4H-SiC spin-qubit based quantum coherent devices with silicon. Optically active defect spin qubits in 4H-SiC, including transition metal ions such as vanadium,[37] and vacancy complexes such as the nitrogen-vacancy center (NV)[38] and the divacancy (VV)[39] are widely investigated for quantum computing,[10] networking,[40] and sensing.[41] The spalling process seeks to overcome scalability challenges of these technologies by creating transferable thin films from semiconductor qubit hosts. We characterize the coherence properties of spin qubits subjected to spalling to infer the quality of native spalled films for quantum applications.

To benchmark the performance of spalled 4H-SiC for quantum applications, we study the spin properties of neutral-divacancy defects ($VV^0$) in spalled high purity semi-insulating (HPSI) 4H-SiC. **Figure 6a** shows a continuous-wave optically detected magnetic resonance (ODMR) spectra in the absence of an applied external field. Both samples show a pronounced resonance of PL4 divacancies centered at 1.353 GHz. Photoluminescence data is additionally included in Figure S4. Coherent spin control of the PL4 ensemble in the spalled film and bulk substrate was then performed to compare spin coherence times. In Figure 6b, Rabi oscillations are observed in the film by sweeping the duration of a single microwave pulse. Additionally, a Ramsey pulse sequence is used to characterize the spin $T_2$* as shown in Figure 6c. A $T_2$* = 1.31 ± 0.03 μs was measured in the bulk wafer, consistent with the literature values for $VV^0$ ensembles at this temperature,[39] and a $T_2$* = 1.02 ± 0.03 μs was measured in the film. Similarly, we perform Hahn echo measurements in Figure 6d to characterize the coherence time $T_2$ between the two samples and find a slightly shorter $T_2$ = 79.7 ± 12.2 μs in the film compared to $T_2$ = 117.4 ± 41.8 μs in the bulk sample. The lower film $T_2$* and $T_2$ are attributed to additional dephasing caused by broadening of the ensemble spin resonances. However, these values are sufficiently high for us to regard the film as quasi-bulk, and demonstrate the viability of these spalled films as a defect center qubit host platform. We speculate that the resonance broadening can be caused by unresolved strain created in the film during the spalling process or electric field noise from the surfaces which are on average closer to the excited divacancies in the ~ 30 μm thick film versus the 500 μm thick bulk substrate.[10] Furthermore, roughness of the spalled surface could lead to strained mounting of the film during the measurement, however spatial dependence of the coherence properties of the film was not explored. A future detailed study which correlates coherence properties to microstructural strain in spalled 4H-SiC could provide guidance on how to optimize the quality of spalled films.

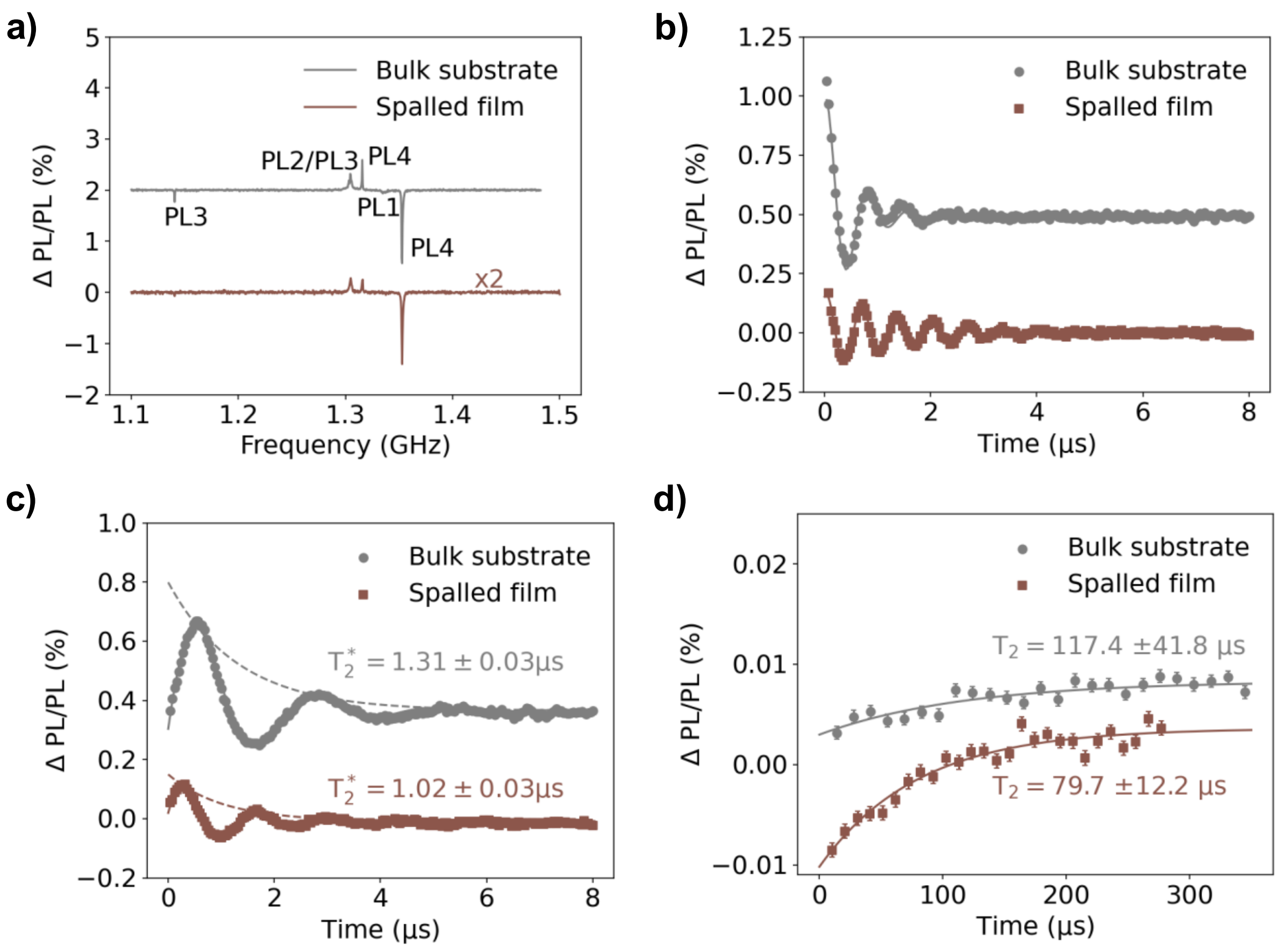

**Figure 6. Comparison of 4H-SiC $VV^0$ qubit properties between bulk wafer and spalled film. a)** Continuous-wave (CW) ODMR spectra for the HPSI bulk wafer and spalled film measured as a function of microwave frequency and detected as a normalized change in photoluminescence intensity. **b)** Rabi oscillations of the basal (PL4) divacancy in HPSI bulk wafer and spalled film. The driving frequency is resonant to the PL4 dip at 1.353 GHz in the CW ODMR or pulsed ODMR spectra. **c)** Ramsey decay of PL4 divacancy defects showing a film $T_2$* ~ 78% of the bulk. The dashed lines illustrate the Ramsey decay envelope. **d)** Hahn echo decay measurements showing a film $T_2$ ~ 68% of the bulk. In b,c,d, solid lines are fits, and error bars are standard errors corresponding to 95% confidence intervals. All measurements are performed at a temperature of 8 K.

## 3. Conclusion

Controlled spalling of 4H-SiC has been successfully demonstrated on a wide variety of substrate types and sizes. Reliable spalling of this high fracture toughness material is made possible by innovations in the stressor layer thickness distribution and spalling crack initiation into the 4H-SiC. Thickness nonuniformity across an entire spalled film is currently ~ 6 μm peak-to-peak, limited primarily by further advancements needed in regulation of the peeling process. Intrinsic roughness of 4-degree miscut substrates can be kept under 1 μm peak-to-peak by spalling perpendicular to the miscut direction. Heterogenous integration and substrate re-use show promise, with more complicated schemes to be pursued in future work. An initial demonstration of coherent spin control of a $VV^0$ ensemble in spalled 4H-SiC yields a spin $T_2^*$ which is similar (~ 78%) to the bulk value and a spin $T_2$ which is likewise similar (~ 68%) to the bulk value, sufficient to motivate future exploration of spalled films incorporated with on-chip silicon photonics. Other goals for future work include spalling full 150 mm and 200 mm wafers of 4H-SiC as well as exploring the spalling of 4H-SiC substrates with prefabricated power devices on the wafer surface.

**Supporting Information**

Methods and Supporting Information available upon reasonable request to the corresponding author.

**Acknowledgements**

S.G. acknowledges support from the Vannevar Bush Fellowship under the program sponsored by the Office of the Undersecretary of Defense for Research and Engineering and in part by the Office of Naval Research as the Executive Manager for the grant (C.P.H., A.W., P.V.K.N., S.G.). This work is also supported by the Air Force Office of Scientific Research under award number FA9550-23-1-0330 (C.P.H., C.Z., D.D.A.). Additional support was provided by the U.S. Department of Energy, Office of Science; Basic Energy Sciences, Materials Sciences, and Engineering Division (C.W., F.J.H.). Work performed at the Center for Nanoscale Materials, a U.S. Department of Energy Office of Science User Facility, was supported by the U.S. DOE, Office of Basic Energy Sciences, under Contract No. DE-AC02-06CH11357. The authors thank Dr. Stephen Bedell for discussions and Dr. Elina Kasman, C. Suzanne Miller, and Adrian Portales for support with experiments.